\documentstyle[aps,preprint,tighten]{revtex}
\newcommand{\news}{\setcounter{equation}{0}}

\def\eqn{\begin{equation}}
\def\eeqn{\end{equation}}
\def\arr{\begin{array}}
\def\earr{\end{array}}
\def\a{\alpha}
\def\s{\sigma}
\def\w{\wedge}
\def\e{\epsilon}
\def\la{\lambda}
\def\z{\zeta}
\font\mybb=msbm10 at 12pt
\def\bb#1{\hbox{\mybb#1}}
\def\bZ {\bb{Z}}
\def\bR {\bb{R}}
\def\bE {\bb{E}}

\def\bM {\bb{M}}

\font\mycc=msbm10 at 10pt
\def\cc#1{\hbox{\mycc#1}}
\def\cZ {\cc{Z}}
\begin{document}
\preprint{DAMTP-R/97/15, UPR-741-T, hep-th/9703204}
\date{March 1997}
\title{Non-threshold D-brane bound states and black holes with non-zero
entropy}
\author{Miguel S. Costa$^1$\footnote{email address: M.S.Costa@damtp.cam.ac.uk}
and Mirjam Cveti\v c$^2$\footnote{email address:
  cvetic@cvetic.hep.upenn.edu}}
\address{$^1${D.A.M.T.P.\\ University of Cambridge, Cambridge, CB3 9EW, UK}
\\ and\\
$^2${Department of Physics and Astronomy
\\
University of Pennsylvania, Philadelphia, PA 19104-6396}}
\maketitle
\begin{abstract}
We start with BPS-saturated configurations of two  (orthogonally) 
intersecting M-branes and  use
the electro-magnetic duality  or  dimensional 
reduction along a boost, in order to obtain new
$p$-brane bound states. 
In the first case the resulting
configurations are interpreted as BPS-saturated non-threshold  bound states of 
intersecting
$p$-branes, and in the second case as $p$-branes intersecting at angles
and their duals. As a by-product we deduce the enhancement 
of supersymmetry as the  angle approaches zero. We also
 comment on the D-brane theory
describing these new bound states, and a connection between the angle and the
world-volume gauge fields of the D-brane system. We use these configurations
to find new embeddings of the four and five dimensional black holes with 
non-zero entropy, whose entropy  now  also depends on the angle and 
world-volume gauge fields. The corresponding D-brane configuration sheds 
light on the microscopic  entropy of such black holes.
\end{abstract}

\newpage

\section{Introduction}
\news

The existence of BPS-saturated configurations of superstring theory
 has become a corner
stone in establishing and testing the different non-perturbative
superstring dualities (\cite{HullTown,Witt} and references there in). The
importance of such BPS-saturated states relies on the fact that  (for $N\ge 4$
supersymmetry) their classical
properties do not receive  quantum corrections and  thus one can 
follow their properties to the strong coupling regime.

As a consequence, over the last year there has been a dramatic progress
in the understanding of different BPS-saturated  $p$-brane solutions in
supergravity theories.
In particular, $p$-brane  configurations
 within $D=11$ supergravity are of special interest, 
 since the latter theory is  the effective field
theory limit of the conjectured M-theory \cite{Witt,Town}. By  
intersecting different $p$-brane configurations, each of them preserving
$1/2$ of the maximal supersymmetry \cite{DuffStel,Guven,Izqu..}, it has
been possible to construct  new solutions preserving a lower fraction of the
maximal supersymmetry
\cite{PapaTown,Tsey,KlebTsey,Bala...,Gaun..,Costa}. These configurations can
be reduced to lower dimensions to obtain supersymmetric $p$-brane
solutions of superstring theories \cite{LuPope,Khvi..,Behr..,Berg..}.
The  BPS-saturated (basic) $p$-brane configurations that intersect 
orthogonally, correspond to  the  marginally bound configurations. 

On the other hand, there exist non-marginally (non-threshold) bound 
BPS-saturated configurations among the composite $p$-branes.
 Typically these configurations interpolate between
different type of configurations. 
To be definite, consider the bound state
of a D-2-brane lying within a D-4-brane as an example. This
configuration can be obtained by reducing the M-theory bound state
interpreted as a membrane lying within a five-brane \cite{Green..},
which in turn arises as a consequence of the $SL(2,\bZ)$
electromagnetic duality of the $D=8$  Type II superstring
theory \cite{HullTown,Izqu..}. The same configuration arises
by performing T-duality at an angle on the Type IIB 3-brane solution
\cite{Brec..,CostaPapa}. Other bound states can be obtained by
reducing a M-theory configuration at an angle or along a boost
\cite{RussoTsey,TseyII}, or by performing a T-duality operation along a boost
\cite{CostaPapa}.

 The aim of this paper is to find new non-threshold
BPS-saturated states of M-theory and superstring 
theories by using these (solution
generating) 
techniques. In particular, we will be interested in finding new D-brane
\cite{Polc} configurations that may be relevant for the black hole
entropy counting \cite{StroVafa,CallMald,Brec..1,MaldStro,John..}.
Therefore most of the solutions presented will be bound states of
D-branes. Their uplift to eleven dimensions  and their
M-theory interpretation should be straightforward. 
We will  also comment on
the D-brane theory associated with the new bound states.

The paper is organised as follows. In Section II we shall study the
action of the $D=8$ type II superstring theory U-duality subgroup
$SL(2,\bZ)$ on the intersecting M-branes. The resulting configurations
will be reduced to D-brane bound states of the  Type IIA theory. 
In Section III, by reducing along a boost intersecting configurations of two
M-branes, we will be able to deduce  the supergravity solutions that
are interpreted as two D-$p$-branes intersecting at an angle
\cite{Berk..,Brec..2}, as well as other $p$-branes intersecting 
at an angle. In both Sections II and III we
will find new embeddings of the supersymmetric $D=4$
\cite{CvetYoum,CvetTsey,CvetTsey1,Hull} and $D=5$
\cite{Tsey1} black holes with non-zero entropy.
In section IV we shall give some concluding remarks. 

Before proceeding, we first settle our notation and conventions. We will be
representing the orthogonal intersection of two configurations $A$ and $B$
by $A\perp B$, and a (non-threshold) configuration interpolating
between $A$ and $B$ by $(A|B)$. For example, in this notation $(2|5)$
represents a membrane lying within a five-brane and $(2|5)\perp (2|5)$
the orthogonal intersection of two of these bound states. We will work
with the following action
for the bosonic sector of  $D=11$ supergravity 
\eqn
I_{11}=\int
d^{11}x\sqrt{-g}\left[R-\frac{1}{2.4!}{{\cal F}}^{2}\right]+
\frac{1}{6}\int {\cal F}\wedge {\cal F}\wedge {\cal A}\ ,
\eeqn
where ${\cal F}=d{\cal A}$ and ${\cal A}$ is a 3-form field. The
$p$-brane supergravity background fields are usually expressed in
terms of harmonic functions on the $d$-dimensional overall transverse
space of the configuration $\bE^d$. For simplicity, we will consider the
case where the
harmonic functions are centered at the origin, i.e.
\eqn
H_i=1+\frac{\a_i}{|{\bf x}|^{d-2}}\ ,
\eeqn
where ${\bf x}$ are the coordinates on $\bE^d$.
All the solutions will be presented in terms of the corresponding
harmonic functions, therefore the multi-centered case follows
immediately.

The electric charges are defined by
\eqn
Q_i=\frac{1}{(d-2)A_{d-1}V_i}\int_{\Sigma_i}\star {\cal F}\ ,
\eeqn
where $A_{d-1}$ is the unit $(d-1)$-sphere volume, $V_i$ is the volume of
the compact space along which the
corresponding membrane is delocalized and ${\Sigma_i}$ is
the product of this compact space with the asymptotic
$(d-1)$-sphere. Similarly, the magnetic charges are defined by
\eqn  
P_i=\frac{1}{(d-2)A_{d-1}V_i}\int_{\Sigma_i} {\cal F}\ ,
\eeqn
where $V_i$ is the volume of the compact space along which the
corresponding five-brane is delocalized and
${\Sigma_i}$ is 
the product of this compact space with the asymptotic
$(d-1)$-sphere. Similar definitions hold for the electric and magnetic
charges associated with a given $n$-form field strength of the lower
dimensional theories.

\section{Electromagnetic duality and bound states of intersecting branes}
\news

Reduction of the M-theory on a 3-torus yields the $D=8$ Type II
superstring theory. The corresponding U-duality group is
$SL(3,\bZ)\times SL(2,\bZ)$ \cite{HullTown}. In this section we will
study the action of the $SL(2,\bZ)$ electro-magnetic duality group
on the intersecting M-brane configurations. Following \cite{Izqu..} we
will start by presenting the action of the $SL(2,\bR)$ electro-magnetic
duality group on the bosonic sector of the $D=8$, $N=2$ supergravity. The
latter theory admits a consistent truncation with the following
bosonic field content: a metric $g_{ab}$, a scalar $\s$, a pseudo
scalar $\rho$ and a 3-form gauge potential ${\cal A}_3$. This
truncation can be embedded in the $D=11$ supergravity by performing
the Ka\l u\.{z}a-Klein Ans\"atze
\eqn
\arr{rl}
ds_{11}^{2}=&e^{\frac{2}{3}\s}ds_{8}^{2}+e^{-\frac{4}{3}\s}d{\bf u}.
d{\bf u}\ , \\\\
{\cal F}=&{\cal F}_4+d\rho\w du_1\w du_2\w du_3\ ,
\earr
\label{(2.1)}\eeqn
where $u_i$ with $i=1,2,3$ are the coordinates on the 3-torus. Next,
define the following complex scalar field $\la$ and 4-form field
strength ${\cal G}$
\eqn
\arr{rl}
\la =&\rho +ie^{-2\s}\ , \\
{\cal G}=&e^{-2\s}\star {\cal F}_{4}-\rho {\cal F}_{4}\ ,
\earr
\label{(2.2)}\eeqn
where $\star$ is the dual with respect to the eight-dimensional metric
in (\ref{(2.1)}). The electro-magnetic duality acts on these fields by the
$SL(2,\bR)$ transformation
\eqn
\arr{rl}
\la\rightarrow&\frac{a\la +b}{c\la +d}\ ,\\
({\cal F},{\cal G})\rightarrow&({\cal F},{\cal G})
\left(
\arr{cc}
d&-b\\
-c&a
\earr
\right)\ .
\earr
\label{(2.3)}\eeqn 
In the quantum theory the $SL(2,\bR)$ group is broken to $SL(2,\bZ)$,
i.e. $a,b,c,d$ are integers satisfying $ad-bc=1$. In the following 
subsections we will
consider the action of the $U(1)$ subgroup, whose matrix elements are
\eqn
\left(
\arr{cc}
\cos{\z}&\sin{\z}\\
-\sin{\z}&\cos{\z}
\earr
\right)\ ,
\label{(2.4)}\eeqn
on the intersecting M-branes. The full $SL(2,\bZ)$
multiplets can be found by acting with an $SL(2,\bZ)$ transformation on 
the set of configurations obtained by acting with the 
$U(1)$ transformation (\ref{(2.4)}) on the intersecting 
M-branes we are considering \cite{Izqu..}. 
Since this $SL(2,\bZ)$  transformation  
changes  only the asymptotic values  of the fields,
we will  restrict ourselves to  configurations which are obtained by 
$U(1)$ transformations only, i.e. we  confine ourselves to
the study of configurations with the  canonical  asymptotic values of 
the  scalar fields.

\subsection{The $(5\perp 5|2\perp 2)$ configuration}

Let us start by considering the $5\perp 5$ M-theory solution
\cite{PapaTown,Tsey}
\eqn
\arr{rl}
ds_{11}^{2}=&(H_1H_2)^{\frac{2}{3}}\left[ 
(H_1H_2)^{-1}\left( -dt^{2}+d{\bf u}.d{\bf u}\right)\right.\\\\
&\left.
+H_1^{-1}ds^2(\bE_1^2)+H_2^{-1}ds^2(\bE_2^2)+ds^2(\bE^3)\right]\ ,\\\\
{\cal F}=&-\star dH_1\w\e(\bE_2^2)-\star dH_2\w\e(\bE_1^2)\ ,
\earr
\label{(2.5)}\eeqn
where $\e(\bE_i^2)$ is the volume form on the corresponding
two-dimensional Euclidean space and $\star$ is the dual with respect
to the Euclidean metric in $\bE^3$. The functions $H_i$ are
harmonic functions on the overall transverse space
$\bE^3$ as described at the end of the introduction. 
Next, perform a $U(1)$ electro-magnetic duality transformation
by using the formulae  given in the  previous subsection. 
The resulting configuration
interpolates between the original $5\perp 5$ and the $2\perp 2$
configurations. In our notation it is represented by $(5\perp 5|2\perp
2)$. The corresponding background fields are given by
\eqn
\arr{rl}
ds_{11}^{2}=&(\tilde{H}_{12}H_1H_2)^{\frac{1}{3}}\left[ 
-(H_1H_2)^{-1}dt^{2}+\tilde{H}_{12}^{-1}ds^2(\bE_1^3)\right.\\\\
&\left.
+H_1^{-1}ds^2(\bE_1^2)+H_2^{-1}ds^2(\bE_2^2)+ds^2(\bE_2^3)\right]\ ,\\\\
{\cal F}=&
-\cos{\z}\left[\star dH_1\w\e(\bE_2^2)+\star dH_2\w\e(\bE_1^2)\right]\\\\
&-\sin{\z}\left[dH_1^{-1}\w\e(\bM_1^3)+dH_2^{-1}\w\e(\bM_2^3)\right]\\\\
&+\cos{\z}\sin{\z}\ d\left(\frac{H_1H_2-1}{\tilde{H}_{12}}\right)\w\e(\bE_1^3)\ ,
\earr
\label{(2.6)}\eeqn
where $\tilde{H}_{12}=\sin^2{\z}+\cos^2{\z}\ H_1H_2$ and $\e(\bM_i^3)$
is the volume form on the Minkowski space $\bR\times\bE_i^2$. If
$\sin{\z}=0$ we obtain the $5\perp 5$ solution and if $\cos{\z}=0$ the
$2\perp 2$ solution. 

The ADM mass density of this composite
configuration can be easily calculated \cite{Lu}. The result is
\eqn
\frac{M}{A_2}=\sqrt{P_1^2+Q_1^2}+\sqrt{P_2^2+Q_2^2}\ ,
\label{(2.7)}\eeqn
where $Q_i$ and $P_i$ are, respectively,  the electric and magnetic
charges of the  constituent M-branes and are given by
\eqn
P_i=\cos{\z}\ \a_i\ ,\ \ \  Q_i=\sin{\z}\ \a_i\ ,\ \ \ i=(1,2).
\label{(2.8)}\eeqn
  
Consider now the reduction on $S^1$ of the solution (\ref{(2.6)}) along one
direction of $\bE_1^3$. It is straightforward to obtain the
corresponding  Type IIA theory background fields. This solution describes a new
bound state of D-branes, and it is represented in our notation as
$(4\perp 4|2\perp 2)$. It interpolates between the $4\perp 4$ D-brane
configuration (say, $n_1$ D-4-branes at the origin lying along the
$Y^1,Y^2,Y^3,Y^4$ directions and $n_2$ D-4-branes at the origin lying
along the $Y^1,Y^2,Y^5,Y^6$ directions) and the $2\perp 2$ D-brane
configuration ($n_1$ D-2-branes at the origin lying along the
$Y^3,Y^4$ directions and $n_2$ D-2-branes at the origin lying
along the $Y^5,Y^6$ directions). As mentioned in \cite{CostaPapa} such
 configurations can be obtained by performing a T-duality at an
angle on orthogonally intersecting configurations.\footnote{This result is
  not surprising since within superstring theory the $SL(2,\cZ)$
  electro-magnetic duality group is a perturbative symmetry 
  \cite{HullTown}.}
In fact, consider
the  Type IIB $3\perp 3$ D-brane configuration ($n_1$ D-3-branes at the
origin lying along the $Y^1,Y^3,Y^4$ directions and $n_2$ D-3-branes at
the origin lying along the $Y^1,Y^5,Y^6$ directions) and rotate the 
configuration in the $Y^1,Y^2$ plane. This D-brane configuration can be
presented as  parallel to the following directions:
\eqn
\arr{rl}
3_1:&Y^3,Y^4,Y^2=\tan{\z}\ Y^1\ ,\\
3_2:&Y^5,Y^6,Y^2=\tan{\z}\ Y^1\ .
\earr
\label{(2.9)}\eeqn
Taking the $Y^1$ and $Y^2$ directions to be circles of radius 
$R_1$ and $R_2$, the angle $\zeta$ obeys the quantisation condition
$\tan{\zeta}=\frac{\omega^2}{\omega^1}\frac{R_2}{R_1}$. Thus, the
D-3-branes are wrapped along the $(\omega^1,\omega^2)$ cycle of the
$Y^1Y^2$ 2-torus. Both D-3-branes in this configuration can be seen
as lying along the $Y^1$ direction but with the $U(1)$ center of the
$U(n_i)$ worldvolume scalar field $Y^2$ turned on as in (\ref{(2.9)}) 
\cite{Witt1}.\footnote{In the cases where the winding number $\omega^1$
is bigger than one it is necessary to introduce extra $U(\omega^1)$ indices 
for the world-volume fields \cite{HashTayl}.}
Similar comments will apply to other configurations presented in
this paper.

Applying T-duality along the $Y^2$ 
direction \cite{Polc1} of the configuration (\ref{(2.9)}) we obtain 
the D-brane theory corresponding to the
$(4\perp 4|2 \perp 2)$ bound
state
\eqn
\arr{rl}
4_1:&Y^1,Y^2,Y^3,Y^4\ ,\\
    &F_{12}=\tan{\z}\ ,\\
4_2:&Y^1,Y^2,Y^5,Y^6\ ,\\
    &F_{12}=\tan{\z}\ ,
\earr
\label{(2.10)}\eeqn
where $F$ is the world-volume 2-form gauge field strength and we have
considered its $U(1)$ factor in the decomposition $U(n_i)=\left(U(1)\times
SU(n_i)\right)/Z_{n_i}$. When this field  is turned on as in (\ref{(2.10)}),  
it ensures the coupling of a D-$(p+2)$-brane to a D-$p$-brane \cite{Doug}. 
Note that it is essential that both
world-volume gauge fields are turned on as in (\ref{(2.10)}). In fact, had we
turned on just the world-volume gauge field associated with the first
group of D-4-branes the corresponding configuration would have been 
interpreted as $(4|2)\perp 4$ in a manner incompatible with the
D-brane intersecting rules \cite{Polc1}. The configuration (\ref{(2.10)})
provides another
example of how the world-volume gauge field is essential for 
 ensuring  supersymmetry  of the configuration \cite{BalaLeig}. 
More generally, by
applying T-duality along transverse directions to the configuration
(\ref{(2.9)}), we have the non-threshold D-brane bound states
\eqn
\biggl( (p+2)\perp (p+2)|p\perp p\biggr)\ ,
\label{(2.11)}\eeqn
for $2\le p\le 5$.

\subsection{The $(2\perp 5|5\perp 2)$ configuration}

For the sake of completeness, we will now repeat the previous analysis for
 the $2\perp 5$
configuration of M-theory \cite{Tsey}
\eqn
\arr{rl}
ds_{11}^{2}=&H_1^{\frac{1}{3}}H_2^{\frac{2}{3}}\left[ 
(H_1H_2)^{-1}ds^2(\bM^2)+H_1^{-1}dy_2^2\right. \\\\
&\left. +H_2^{-1}\left(dy_3^2+d{\bf u}.d{\bf u}\right)+
  ds^2(\bE^4)\right]\ ,\\\\
{\cal F}=&-dH_1^{-1}\w\e(\bM^2)\w dy_2-\star dH_2\w dy_2\ ,
\earr
\label{(2.12)}\eeqn
where we use the same notation as in (\ref{(2.5)}). Performing a $U(1)$
electro-magnetic duality transformation as in the previous subsection
we obtain the following solution
\eqn
\arr{rl}
ds_{11}^{2}=&\left(\tilde{H}_{12}H_1H_2\right)^{\frac{1}{3}}\left[ 
(H_1H_2)^{-1}ds^2(\bM^2)+H_1^{-1}dy_2^2\right. \\\\
&\left. +H_2^{-1}dy_3^2+\tilde{H}_{12}^{-1}ds^2(\bE^3) +
  ds^2(\bE^4)\right]\ ,\\\\
{\cal F}=&-\cos{\z}\left[dH_1^{-1}\w\e(\bM^2)\w dy_2 + \star
  dH_2\w dy_2\right]\\\\ 
&+\sin{\z}\left[ \star dH_1\w dy_3 +dH_2^{-1}\w\e(\bM^2)\w dy_3 \right]\\\\
&+\cos{\z}\sin{\z}\ d\left(\frac{H_2 -H_1}{\tilde{H}_{12}}\right)\w\e(\bE^3)\ ,
\earr
\label{(2.13)}\eeqn
where $\tilde{H}_{12}=\sin^2{\z}\ H_1 +\cos^2{\z}\ H_2$. In both limiting
cases, $\sin{\z}=0$ and $\cos{\z}=0$, this solution  reduces to
the $2\perp 5$ solution. However, the constituent M-branes are
located along different directions. Also, as $\z$ varies the electric
and magnetic charges are interchanged. The ADM mass density of our
solution is again given (up to a constant) by (\ref{(2.7)}). However, the 
electromagnetic charges are now given by
\eqn
\arr{l}
Q_1=\cos{\z}\ \a_1\ ,\ \ \  P_1=-\sin{\z}\
\a_1\ ,\\\\ 
P_2=\cos{\z}\ \a_2\ ,\ \ \  Q_2=\sin{\z}\
\a_2\ .
\earr
\label{(2.14)}\eeqn

We proceed by describing the  Type IIA D-brane bound state associated with
the solution (\ref{(2.13)}). Reducing along one direction of $\bE^3$ we get
the D-brane bound state $(2\perp 4|4\perp 2)$. It interpolates between
the $2\perp 4$ configuration (say, $n_1$ D-2-branes lying
along the $Y^1,Y^2$ directions and $n_2$ D-4-branes lying
along the $Y^1,Y^3,Y^4,Y^5$ directions) and the $4\perp 2$ D-brane
configuration ($n_1$ D-4-branes lying along the
$Y^1,Y^2,Y^4,Y^5$ directions and $n_2$ D-2-branes lying
along the $Y^1,Y^3$ directions). Again, we can derive this
configuration by performing T-duality at an angle on the  Type IIB $3\perp 3$
D-brane configuration. Starting with $n_1$ D-3-branes
lying along the $Y^1,Y^2,Y^4$ directions and $n_2$ D-3-branes lying
along the $Y^1,Y^3,Y^5$ directions and rotating the 
configuration in the $Y^4Y^5$ plane we have
\eqn
\arr{rl}
3_1:&Y^1,Y^2,Y^4=\cot{\z}\ Y^5\ ,\\
3_2:&Y^1,Y^3,Y^4=-\tan{\z}\ Y^5\ .
\earr
\label{(2.15)}\eeqn
Applying T-duality along $Y^4$ we obtain the D-brane theory
corresponding to the $(2\perp 4|4\perp 2)$ bound state
\eqn
\arr{rl}
4_1:&Y^1,Y^2,Y^4,Y^5\ ,\\
    &F_{45}=-\cot{\z}\ ,\\
4_2:&Y^1,Y^3,Y^4,Y^5\ ,\\
    &F_{45}=\tan{\z}\ .
\earr
\label{(2.16)}\eeqn
Once again, the  turned on world-volume gauge fields in  (\ref{(2.16)}) 
ensure that the  configuration remains supersymmetric.
More generally,  there are the non-threshold bound states of
D-branes
\eqn
\biggl( p\perp (p+2)|(p+2)\perp p\biggr)\ ,
\label{(2.17)}\eeqn
for $1\le p\le 6$.

\subsection{Black holes with non-zero entropy}

We proceed by presenting new embeddings of the supersymmetric black
holes with non-zero entropy. Consider the solution
described by  (\ref{(2.13)}). There is a direction common
to all the constituent M-branes. Therefore we can add a plane wave
along this direction \cite{Tsey}. The resulting solution is described
by the metric
\eqn
\arr{rl}
ds_{11}^{2}=&\left(\tilde{H}_{12}H_1H_2\right)^{\frac{1}{3}}\left[ 
(H_1H_2)^{-1}\left( -dt^2 +dy_1^2 +(H_3-1)(dy_1\pm dt)^2\right)\right. \\\\
&\left. +H_1^{-1}dy_2^2+H_2^{-1}dy_3^2+\tilde{H}_{12}^{-1}ds^2(\bE^3) +
  ds^2(\bE^4)\right]\ ,
\earr
\label{(2.18)}\eeqn
and the 4-form field strength remains the same. This solution
provides a new embedding of the $D=5$ black hole with non-zero
entropy. By calculating the horizon area $A_H$, we obtain the following
expression for the Bekenstein-Hawking entropy (we have introduced the
five-dimensional Newton constant)
\eqn
S_{BH}=\frac{A_H}{4G^{5}_N}=\frac{A_3}{4G^{5}_N}\sqrt{\a_1\a_2\a_3}=
\frac{A_3}{4G^{5}_N}\sqrt{\left(|Q_1P_2|+|Q_2P_2|\right)|Q_3|},
\label{(2.19)}\eeqn
where $Q_3$ is the electric charge associated
with the Ka\l u\.{z}a-Klein(KK) 
modes along the $y_1$ direction in (\ref{(2.18)}). In the quantum mechanical
D-brane theory the charges $Q_i$ and $P_i$ become quantised (this fact
follows from the quantisation of the angle $\zeta$ and parameters $\alpha_i$), e.g.,
$|Q_3|=c_3q_3$ where $c_3$ is the quantum of charge and $q_3$ a positive
integer. We then expect the entropy to have the following integer
valued form
\eqn
S_{BH}=2\pi\sqrt{\left(q_1p_2+q_2p_2\right)q_3}\ .
\label{new}\eeqn
Compactifying this solution, as in the
previous subsection,  we obtain the D-brane $(2\perp 4|4\perp 2)$ bound
state (\ref{(2.16)}) with an extra momentum along the $Y^1$
direction. Applying T-duality along $Y^2$ we obtain the $(1_D\perp
5_D|3\perp 3)$ bound state\footnote{We use the subscripts D (Dirichlet), 
S (solitonic) and F (fundamental) whenever there is ambiguity, i.e. in 
the Type IIB theory and for $p=1,5$.} described by the following D-branes
\eqn
\arr{rl}
3:&Y^1,Y^4,Y^5\ ,\\
  &F_{45}=-\cot{\z}\ ,\\
5:&Y^1,Y^2,Y^3,Y^4,Y^5\ ,\\
  &F_{45}=\tan{\z}\ ,
\earr
\label{(2.20)}\eeqn
carrying momentum along the $Y^1$ direction. This bound state
interpolates between the D-brane configuration used by Callan and
Maldacena \cite{CallMald} in their black hole entropy counting, i.e.
the $1_D\perp 5_D$
configuration, and the $3\perp 3$ configuration. It would be extremely
interesting to obtain
the entropy formula (\ref{(2.19)}) by using the D-brane counting associated
with the configuration described by (\ref{(2.20)}).

Finally, let us note that by adding a KK monopole \cite{Costa} along
one of the directions of $\bE^4$ in (\ref{(2.18)}) we have an embedding of
the four-dimensional black hole with non-zero entropy.  

Another $D=4$ embedding  can be obtained by performing the 
electromagnetic duality
described at the beginning of this section on the $2\perp 2\perp
5\perp 5$ configuration \cite{KlebTsey}.\footnote{The same solution
 can be obtained by applying T-duality at an angle on the $3\perp
  3\perp 3\perp 3$  Type IIB solution.} This can be done by identifying the
three ${\bf u}$ directions in (\ref{(2.1)}) with the three common directions 
to both five-branes. The resulting entropy is\footnote{Note that this 
solution is  $SL(2,\cZ)$  transformed of the basic four charge  
configurations with  two electric and two magnetic charges \cite{CvetYoum}.} 
\eqn
S_{BH}=\frac{A_2}{4G^{4}_N}\sqrt{\a_1\a_2\a_3\a_4}=
\frac{A_2}{4G^{4}_N}\sqrt{|Q_1Q_2P_3P_4|+|P_1P_2Q_3Q_4|}\ .
\label{(2.21)}\eeqn
where  
\eqn
\arr{l}
Q_i=\cos{\z}\ \a_i\ ,\ \ \  P_i=-\sin{\z}\ 
\a_i\ , \ \ i=(1,2)\ ,\\\\ 
P_i=\cos{\z}\ \a_i\ ,\ \ \  Q_i=\sin{\z}\
\a_i\ , \ \ \ i=(3,4)\ . \\ 
\earr
\label{(2.22)}\eeqn
Charge quantisation should give the entropy an integer valued form as in  
(\ref{new}).

\section{Branes at angles and their duals}
\news

In this section, we start with known solutions, i.e., two
orthogonally intersecting p-branes,  and  use a solution generating
technique to obtain  new solutions that are interpreted 
as configurations of D-$p$-branes (as well as other $p$-branes) 
intersecting at an angle (different from $\pi/2$ or zero).  The existence of
these former configurations has been deduced in \cite{Berk..} by using
D-brane quantum mechanics, and the corresponding supergravity solutions
correspond to those in \cite{Brec..2}. 
Here, we generate such solutions of Type IIA supergravity 
by reducing the $2\perp 2$ and $2\perp 5$  configuration of the 
M-theory along a specific boost direction and further duality 
transformations. We will find new T-duals of the branes intersecting at
an angle and comment on the corresponding D-brane theory.

\subsection{D-$p$-branes intersecting at an angle}

Let us start with the $2\perp 2$ M-brane configuration
\cite{PapaTown,Tsey}
\eqn
\arr{rl}
ds_{11}^{2}=&(\tilde{H}_1\tilde{H}_2)^{\frac{1}{3}}\left[ 
-(\tilde{H}_1\tilde{H}_2)^{-1}dt^{2}+\tilde{H}_1^{-1}ds^2(\bE_1^2)\right.\\\\
&\left.
+\tilde{H}_2^{-1}ds^2(\bE_2^2)+du^2+ds^2(\bE^5)\right]\ ,\\\\
{\cal F}=&-d(\tilde{H}_1^{-1}-1)\w\e(\bM_1^3)-
d(\tilde{H}_2^{-1}-1)\w\e(\bM_2^3)\ ,
\earr
\label{(3.1)}\eeqn
where  we conveniently parameterise 
$\tilde{H}_i=\sin^2{\z}+\cos^2{\z}\ H_i$ and $H_i$ are harmonic
functions on the Euclidean space $\bE^5$. Next, perform the
following   coordinate  (boost) transformation\cite{RussoTsey}:
\eqn
\arr{rl}
t\rightarrow&\cos^{-1}{\z}\left( t+\sin{\z}\ u\right)\ ,\\
u\rightarrow&\cos^{-1}{\z}\left( u+\sin{\z}\ t\right)\ ,
\earr
\label{(3.2)}\eeqn
where we conveniently relate  the boost parameter  $\beta$
 to the  parameter  $\z$  (in the modified harmonic functions $\tilde H_i$), as
 $\cosh\beta=\cos^{-1}\z$.
 Reducing along the $u$ direction yields a {\it static} configuration 
of Type IIA  theory:
\eqn
\arr{rl}
ds_{10}^2=&H_{12}^{\frac{1}{2}}
\left[ -H_{12}^{-1}dt^2+\tilde{H}_1^{-1}ds^2(\bE_1^2)
+\tilde{H}_2^{-1}ds^2(\bE_2^2)+ds^2(\bE^5)\right]\ ,\\\\
e^{2\phi}=&
\left(\tilde{H}_1\tilde{H}_2\right)^{-1}H_{12}^{\frac{3}{2}}\ ,\\\\
{\cal F}_4=&-\cos^{-1}{\z}\left[ d(\tilde{H}_1^{-1}-1)\w\e(\bM_1^3)+
d(\tilde{H}_2^{-1}-1)\w\e(\bM_2^3)\right]\ ,\\\\
{\cal H}=&-\tan{\z}\left[ d(\tilde{H}_1^{-1}-1)\w\e(\bE_1^2)+
d(\tilde{H}_2^{-1}-1)\w\e(\bE_2^2)\right]\ ,\\\\
{\cal F}_2=&-\sin{\z}\ d\left( H_{12}^{-1}-1\right)\w dt\ ,
\earr
\label{(3.3)}\eeqn
where $H_{12}$ is defined by
$\tilde{H}_1\tilde{H}_2=\sin^2{\z}+\cos^2{\z}\ H_{12}$. This solution
interpolates between the $1/4$ supersymmetric $2\perp 2$ solution
(for $\sin{\z}=0$) and the $1/2$ supersymmetric D-0-brane solution (for
$\cos{\z}=0$). In our notation it is expressed as $(2\perp 2|0)$.

The ADM mass density can be expressed in terms of the
D-2-brane charges $Q_1$ and $Q_2$, and the D-0-brane charge $Q_0$ 
as\footnote{We remark that, in this case, it is not necessary to take the 
modulus of $Q_1$ and $Q_2$. However, if we had started with 2-branes 
with opposite charge in (\ref{(3.1)}) that would be the case.}
\eqn
\frac{M}{3A_4}=\sqrt{\left(|Q_1|+|Q_2|\right)^2+Q_0^2}\ ,
\label{(3.4)}\eeqn
where
\eqn
Q_i=\cos{\z}\ \a_i\ ,\ \ \ Q_0=\sin{\z}\
\left(\a_1+\a_2\right)\ , \ \ i=(1,2). 
\label{(3.5)}\eeqn

More generally, by applying T-duality \cite{Berg..1} along the overall
transverse space of the solution (\ref{(3.3)})  we obtain the D-brane 
non-threshold bound states
\eqn
\biggl( p\perp p|p-2\biggr)\ ,
\label{(3.6)}\eeqn
for $2\le p\le 7$. We will comment on the D-brane theory describing
this configuration later.

In order to obtain the supergravity solution interpreted as D-2-branes
intersecting at an angle we perform a T-duality transformation
 along
one direction  of each D-2-brane in the configuration described by
(\ref{(3.3)}), say along $y_1$ and $y_3$. The resulting solution is
\eqn
\arr{rl}
ds_{10}^2=&H_{12}^{-\frac{1}{2}}\left[-dt^2
+\tilde{H}_1dy_1^2
+2\sin{\z}\cos{\z}\ (1-H_1)dy_1dy_2\right.\\\\
&+\tilde{H}_1^{-1}
\left(H_{12}+\sin^2{\z}\cos^2{\z}\ (1-H_1)^2\right)dy_2^2\\\\
&+\tilde{H}_2dy_3^2+2\sin{\z}\cos{\z}\ (1-H_2)dy_3dy_4\\\\
&\left.+\tilde{H}_2^{-1}
\left(H_{12}+\sin^2{\z}\cos^2{\z}\ (1-H_2)^2\right) dy_4^2\right]
+H_{12}^{\frac{1}{2}}ds^2(\bE^5)\ ,\\\\
e^{2\phi}=&H_{12}^{\frac{1}{2}}\ ,\\\\
{\cal A}_3=&-\cos{\z}\ H_{12}^{-1}dt\w\left[\tilde{H}_2(H_1-1)dy_2\w dy_3
+\tilde{H}_1(H_2-1)dy_1\w dy_4\right]\\\\
&+\sin{\z}\ \left( H_{12}^{-1}-1\right) dt\w dy_1\w dy_3\\\\
&+\sin{\z}\cos^2{\z}\ H_{12}^{-1}(H_1-1)(H_2-1)dt\w dy_2\w
dy_4\ .
\label{(3.7)}\earr
\eeqn
Note that the off-diagonal components of this metric arise, after the
T-duality transformations, from the NS-NS 2-form gauge potential
${\cal B}$ in  (\ref{(3.3)}). We claim that this solution should be interpreted
as the intersection at an angle of two sets of parallel D-2-branes.
In fact, taking $H_2=1$ (and therefore $\tilde{H}_2=1$
and $H_{12}=H_1$) this solution is interpreted as D-2-branes along $y_3$
and making an angle $\frac{\pi}{2}-\z$ with the $y_1$ direction in the
$y_1y_2$ plane. Similarly, taking $H_1=1$ we have D-2-branes along $y_1$
and making an angle $\frac{\pi}{2}-\z$ with the $y_3$ direction in the
$y_3y_4$ plane. 
 
This configuration constitutes a special case of the type of solutions 
presented in \cite{Brec..2}. To be more precise, after a rotation in the
$y_1y_2$ plane we were able to obtain the solution (20) of 
\cite{Brec..2}.\footnote{Note that a different kind of configuration 
interpreted as D-branes intersecting at an angle bounded to a fundamental
string was obtained in \cite{BehrCvet}. There, in general, the metric cannot be diagonalized by a constant $SO(2)$ rotation.}
If the harmonic functions are centered at the same point, the metric in
(\ref{(3.7)}) can be diagonalized and this solution is seen to be the
standard $2\perp 2$ D-brane solution \cite{MichMyers}.
The  configuration at angles can  thus be brought into an orthogonal 
one by  a constant $SO(2)$ rotation (which is  part of the $U$-duality  symmetry group). 

It is interesting that we have been able to deduce the existence of
this configuration by starting with the $2\perp 2$ M-brane configuration
and by following a duality orbit. In the spirit of \cite{CostaPapa}, the
existence of branes at angles may be seen as a consequence of
diffeomorphism invariance of the underlying theories.

Importantly, for $\z\ne \pi/2$ the configuration preserves 1/4
of supersymmetry; for the special case  $\sin{\z}=0$ it is the
$2\perp 2$ D-brane configuration.  On the other hand, for  $\z=\pi/2$ (i.e. 
$\cos{\z}= 0$)  the angle between the intersecting D-2-branes
is zero and the number of preserved 
supersymmetries is increased.  In this limit the  configuration is the 1/2
supersymmetric D-2-brane configuration.\footnote{
 From the D-brane world-volume approach the enlargement of
 supersymmetry as the angle between D-branes approaches zero and the role 
 of the world-volume gauge fields in this process, was studied  
in \cite{BalaLeig}.}

The ADM mass density is given by
\eqn
M=\frac{1}{3A_4}\left(\a_1+\a_2\right)\ ,
\label{(3.8)}\eeqn
which is the sum of the constituent D-2-brane masses. Therefore this is a
D-brane bound state at threshold. Defining the electric
charges \cite{Brec..2}
\eqn
Q_{ij}=\frac{1}{3A_4V_{ij}}\int_{\Sigma_{ij}}\star{\cal F}\ ,
\label{(3.9)}\eeqn
where the integration is over the asymptotic 4-sphere and the
compactified $y_iy_j$ space with volume $V_{ij}$, we have
\eqn
Q_{24}=\sin{\z}\left(\a_1+\a_2\right)\ ,\ \ \ 
Q_{14}=-\cos{\z}\ \a_1\ ,\ \ \ 
Q_{23}=-\cos{\z}\ \a_2\ .
\label{(3.10)}\eeqn
In terms of these charges the ADM mass can be written as
\eqn
\frac{M}{3A_4}=\sqrt{\left(|Q_{14}|+|Q_{23}|\right)^2+Q_{24}^2}\ .
\label{(3.11)}\eeqn
Of course applying T-duality along the overall transverse directions
in (\ref{(3.7)}) we obtain solutions interpreted as D-$p$-branes intersecting
at an angle for $2\le p\le 7$ .

Let us now comment on the D-brane configuration $(2\perp 2|0)$
described by the solution  (\ref{(3.3)}). It is related to the D-2-branes
intersection at an angle  (Eq. (\ref{(3.7)})) by T-duality transformations along
the $y_1$ and $y_3$ directions. The latter configuration can be
represented as
\eqn
\arr{rl}
2_1:&Y^3,Y^1=\tan{\z}\ Y^2\ ,\\
2_2:&Y^1,Y^3=\tan{\z}\ Y^4\ .
\earr
\label{(3.12)}\eeqn
Applying T-duality along the $Y^1$ and $Y^3$ directions we obtain the
D-brane theory describing the $(2\perp 2|0)$ bound state
\eqn
\arr{rl}
2_1:&Y^1,Y^2\ ,\\
&F_{12}=-\tan{\z}\ ,\\
2_2:&Y^3,Y^4\ ,\\
&F_{34}=-\tan{\z}\ .
\earr
\label{(3.13)}\eeqn
Note that if one of the world-volume 2-form gauge fields, associated
with a given set of D-2-branes, were not turned on the corresponding
configuration would not be supersymmetric.

Next, we rewrite the D-2-brane configuration (\ref{(3.12)})
 in the following way
\eqn
\arr{rl}
2_1:&Y^3,Y^2=\cot{\z}\ Y^1\ ,\\
2_2:&Y^1,Y^4=\cot{\z}\ Y^3\ ,
\earr
\label{(3.14)}\eeqn
and apply T-duality along the $Y^2$ and $Y^4$ directions. The
resulting configuration interpolates between the $1/4$ supersymmetric
$2\perp 2$ D-brane configuration and the $1/2$ supersymmetric
D-4-brane configuration, as can be seen by studying the corresponding
supergravity solution (the off-diagonal terms in the metric in (\ref{(3.7)})
will be traded for the NS-NS 2-form gauge potential). This $(2\perp
2|4)$ D-brane bound state is then described by
\eqn
\arr{rl}
4_1:&Y^1,Y^2,Y^3,Y^4\ ,\\
&F_{12}=\cot{\z}\ ,\\
4_2:&Y^1,Y^2,Y^3,Y^4\ ,\\
&F_{34}=\cot{\z}\ .
\earr
\label{(3.15)}\eeqn
We remark that this is not the same bound state found by Lifschytz
\cite{Lifs} (without the extra D-0-brane) whose  classical 
$1/2$ supersymmetric  counterpart solution has been found in
\cite{Brec..}. Such a configuration
couples to a D-0-brane due to the coupling term 
$tr\left(\int F\w F\w {\cal A}\right)$ in the D-4-brane world-volume
action, which in turn requires that the
$F_{21}$ and $F_{43}$ components of the world-volume 2-form gauge field
strength are associated with the same set of D-4-branes as
opposed to our configuration. 

Of course the above set of  solutions generalise to the
\eqn
\biggl( p\perp p|(p+2)\biggr)
\label{(3.16)}\eeqn
non-threshold D-brane bound states for $2\le p\le 6$.

\subsection{Other $p$-branes intersecting at an angle}

We now consider the reduction along a boost of the $2\perp 5$ M-brane
solution (\ref{(2.12)}) that we rewrite in the following way
\eqn
\arr{rl}
ds_{11}^{2}=&\tilde{H}_1^{\frac{1}{3}}\tilde{H}_2^{\frac{2}{3}}\left[ 
\left(\tilde{H}_1\tilde{H}_2\right)^{-1}ds^2(\bM^2)+
\tilde{H}_1^{-1}dy_2^2\right.\\\\ 
&\left.
+\tilde{H}_2^{-1}ds^2(\bE^4)+du^2+ds^2(\bE^3)\right]\ ,\\\\
{\cal F}=& -d\left(\tilde{H}_1^{-1}-1\right)\w\e(\bM^2)\w dy_2
-\star d\tilde{H}_2\w dy_2\w du\ .
\earr
\label{(3.17)}\eeqn
Performing the coordinate transformation (\ref{(3.2)}) and reducing along the
$u$ direction we obtain the  Type IIA solution
\eqn
\arr{rl}
ds_{10}^2=&\left(\tilde{H}_2 H_{12}\right)^{\frac{1}{2}}\left[
-H_{12}^{-1}dt^{2}+\left(\tilde{H}_1 \tilde{H}_2\right)^{-1}dy_1^2\right.\\\\
&\left.
+\tilde{H}_1^{-1}dy_2^2+\tilde{H}_2^{-1}ds^2(\bE^4)+ds^2(\bE^3)\right]\
,\\\\
e^{2\phi}=&
\tilde{H}_1^{-1}\tilde{H}_2^{-\frac{1}{2}}H_{12}^{\frac{3}{2}}\ ,\ \ \
\ {\cal F}_2=-\sin{\z}\ d\left( H_{12}^{-1}-1\right)\w dt\ ,\\\\
{\cal F}_4=&
-\cos^{-1}{\z}\ d\left(\tilde{H}_1^{-1}-1\right)\w dt\w dy_1\w dy_2\\\\
&-\sin{\z}\cos{\z}\ \star dH_2\w dy_2\w dt\ ,\\\\
{\cal H}=&
-\cos{\z}\ \star dH_2\w dy_2 -\tan{\z}\ d\left(\tilde{H}_1^{-1}-1\right)\w
dy_1\w dy_2\ ,
\earr
\label{(3.18)}\eeqn
where $H_{12}$ is defined as in (\ref{(3.3)}). This solution interpolates
between the $2\perp 5$ solution (for $\sin{\z}=0$) and the D-0-brane
solution (for $\cos{\z}=0$), i.e. it is the $(2\perp 5|0)$ bound
state.

Next, apply T-duality along the $y_1$ direction. The resulting Type IIB
solution is
\eqn
\arr{rl}
ds_{10}^2=&\tilde{H}_2^{\frac{1}{2}}H_{12}^{-\frac{1}{2}}\left[
-dt^2+\tilde{H}_1dy_1^2
+2\sin{\z}\cos{\z}\ (1-H_1)dy_1dy_2\right.\\\\
&\left.+\tilde{H}_1^{-1}
\left(H_{12}+\sin^2{\z}\cos^2{\z}\ (1-H_1)^2\right)dy_2^2\right]\\\\
&+ \tilde{H}_2^{\frac{1}{2}}H_{12}^{\frac{1}{2}}\left[
  \tilde{H}_2^{-1}ds^2(\bE^4)+ds^2(\bE^3)\right]\ ,\\\\
e^{2\phi}=&H_{12}\ ,\ \ \ \ {\cal B}=-\cos{\z}\ \omega\w dy_2\ ,\\\\
{\cal A}_4=&-\sin{\z}\cos{\z}\ H_{12}^{-1}\ \omega\w dt\w dy_1\w dy_2\\\\
&-\tan{\z}\ \left(\tilde{H}_2^{-1}-1\right)\w \e(\bE^4)\ ,\\\\
{\cal A}_2=&\sin{\z}\ \left(H_{12}^{-1}-1\right)dt\w dy_1
-\cos{\z}\ \tilde{H}_2 H_{12}^{-1}\left(H_1-1\right) dt\w dy_2\ ,
\earr
\label{(3.19)}\eeqn
where $d\omega=\star dH_2$ and $\star$ is the dual with respect to the
Euclidean metric in $\bE^3$. Note that the second term in ${\cal A}_4$
arises by imposing the self-duality condition ${\cal F}_5=\star {\cal F}_5$ by
hand, where now $\star$ is the dual with respect to the
ten-dimensional metric. 

This solution should be
interpreted as a D-string intersecting the $(5_S|1_D)$ bound state  at an
angle. In fact, taking $H_2=1$ (and therefore $\tilde{H}_2=1$
and $H_{12}=H_1$) this solution corresponds to a D-string in the
$y_1y_2$ plane making an angle $\frac{\pi}{2}-\z$ with the $y_1$
direction. Similarly, taking $H_1=1$ we have the bound state of a
D-string lying within a solitonic 5-brane, i.e. the $(5_S|1_D)$ bound
state. If $\sin{\z}=0$ we have the $1_D\perp 5_S$ solution and if 
$\cos{\z}=0$ the D-string solution. Note that an S-duality
transformation will generate a configuration interpreted as a
fundamental string intersecting a $(5_D|1_F)$ bound state at an angle.

As in the case of the solution (\ref{(3.7)}), the metric in 
(\ref{(3.19)}) can be diagonalized if the harmonic functions are
centered at the same point. However, in this case this solution is 
different to any standard orthogonally intersecting $p$-brane solution.
Since the constituent $p$-branes are charged with respect to different
fields, they are still distinguished by the supergravity solution
as opposed to the case (\ref{(3.7)}).

Since our main goal is to find bound states of D-branes, we proceed by
performing a T-duality transformation along one of the directions in
$\bE^4$ belonging to the solitonic 5-brane. As expected the
corresponding  Type IIA theory solution is interpreted as a D-2-brane
intersecting a $(5|2)$ bound state at an angle. Next, we lift this
solution to $D=11$ and reduce it along 
a direction  internal to the
5-brane (of M-theory). The resulting solution is interpreted as a D-2-brane
intersecting a $(4|2)$ bound state at an angle. The background fields
are given by
\eqn
\arr{rl}
ds_{10}^2=&H_{12}^{-\frac{1}{2}}\left[
-dt^2+dy_5^2+\tilde{H}_1dy_1^2
+2\sin{\z}\cos{\z}\ (1-H_1)dy_1dy_2\right.\\\\
&\left.+\tilde{H}_1^{-1}
\left(H_{12}+\sin^2{\z}\cos^2{\z}\ (1-H_1)^2\right)dy_2^2\right]\\\\
&+ H_{12}^{\frac{1}{2}}\left[
  \tilde{H}_2^{-1}ds^2(\bE^2)+ds^2(\bE^4)\right]\ ,\\\\
e^{2\phi}=&H_{12}^{\frac{1}{2}}\tilde{H}_2^{-1}\ ,\ \ \ \ 
{\cal B}=\tan{\z}\ \left(\tilde{H}_2^{-1}-1\right)\w \e(\bE^2)\ ,\\\\
{\cal A}_3=&\cos{\z}\ \omega\w dy_2+
\sin{\z}\ \left(H_{12}^{-1}-1\right)dt\w dy_1\w dy_5\\\\
&-\cos{\z}\ \tilde{H}_2 H_{12}^{-1}\left(H_1-1\right) dt\w dy_2\w
dy_5\ ,
\earr
\label{(3.20)}\eeqn
where $d\omega =\star dH_2$ and now $H_i$ are harmonic functions on
$\bE^4$. 

The claimed interpretation for this solution can be seen by taking the
appropriate limits: for $H_2=1$ we have a D-2-brane along the $y_5$
direction and making an angle $\frac{\pi}{2}-\z$ with the $y_1$
direction in the $y_1y_2$ plane; for $H_1=1$ we have the bound state
of a D-2-brane in the $y_1y_5$ plane lying within a D-4-brane in the 
$y_1y_5$ and $\bE^2$ planes. This configuration interpolates between
the $2\perp 4$ and D-2-brane configurations and can be represented as
$(2\perp 4|2)$. As in (\ref{(3.9)}) we can define the electric charges
\eqn
Q_{ijk}=\frac{1}{2A_3V_{ijk}}\int_{\Sigma_{ijk}}\star{\cal F}\ ,
\label{(3.21)}\eeqn
where the integration is over the asymptotic 3-sphere and the
compactified $y_iy_jy_k$ space with volume $V_{ijk}$. With this
definitions the non-zero charges are
\eqn
Q_{234}=\sin{\z}\ (\a_1+\a_2)\ ,\ \ \ 
Q_{134}=-\cos{\z}\ \a_1\ .
\label{(3.22)}\eeqn
The magnetic charge carried by the D-4-brane is
\eqn
P_2=-\cos{\z}\ \a_2\ .
\label{(3.23)}\eeqn
The ADM mass density is then seen to be
\eqn
M=\frac{1}{2A_3}(\a_1+\a_2)=\frac{1}{2A_3}
\sqrt{\left( |P_2|+|Q_{134}|\right)^2+Q_{234}^2}\ ,
\label{(3.24)}\eeqn
which is the sum of the masses of the intersecting branes, i.e. 
the D-2-brane and the $(4|2)$ D-brane bound state.

We now comment on the D-brane theory describing the solution
(\ref{(3.20)}). This can be done by realizing that this D-brane configuration
is in fact T-dual to a configuration of D-3-branes intersecting at an
angle. To be more precise, applying T-duality along $Y^5$ and $Y^3$ on
the configuration (\ref{(3.12)}) we obtain the desired result
\eqn
\arr{rl}
2:&Y^5,Y^1=\tan{\z}\ Y^2\ ,\\
4:&Y^5,Y^1,Y^3,Y^4\ ,\\
&F_{34}=-\tan{\z}\ ,
\earr
\label{(3.25)}\eeqn
which confirms our interpretation for this D-brane bound state.
More generally, applying T-duality along $Y^5$ or along the overall
transverse space directions we can have a D-$p$-brane intersecting a 
$(p+2|p)$ bound state at an angle for $1\le p\le 6$.

A new solution interpreted as a D-4-brane intersecting a $(2|4)$ bound
state at an angle  can be
obtained by performing T-duality along two orthogonal directions of
$\bE^2$ in (\ref{(3.20)}). In fact, such solution is just the electro-magnetic
dual of the solution (\ref{(3.20)}) and could be obtained by using the methods
described in section two. The resulting  configuration
interpolates between
the $4\perp 2$ and the D-4-brane configurations, and can be represented
as $(4\perp 2|4)$. The corresponding D-brane theory is obtained by
performing T-duality transformations along the $Y^3$ and $Y^4$
directions in (\ref{(3.25)})
\eqn
\arr{rl}
4_1:&Y^3,Y^4,Y^5,Y^1=\tan{\z}\ Y^2\ ,\\
4_2:&Y^3,Y^4,Y^5,Y^1\ ,\\
&F_{34}=\cot{\z}\ .
\earr
\label{(3.26)}\eeqn
More generally, applying T-duality along $Y^5$ or along the overall
transverse space directions we can have a D-$p$-brane intersecting a
$(p-2|p)$ bound state at an angle, for $3\le
p\le 8$.

Finally, a T-duality transformation along $Y^2$ and $Y^5$ will
bring the configuration (\ref{(3.25)}) to
\eqn
\arr{rl}
2:&Y^1,Y^2\ ,\\
&F_{12}=\cot{\z}\ ,\\
4:&Y^1,Y^2,Y^3,Y^4\ ,\\
&F_{34}=-\tan{\z}\ ,
\earr
\label{(3.27)}\eeqn
which can be generalised to 
\eqn
\biggl((p-2)\perp (p+2)|p\biggr)
\label{(3.28)}\eeqn
D-brane configuration for $2\le p\le 6$.

\subsection{More black holes with non-zero entropy}

As in section two,  we use the above set of configurations
at angles in order to obtain  new embeddings of
the supersymmetric black holes with non-zero entropy. This can be done
by realizing that both  D-brane configurations (\ref{(3.25)}) and 
(\ref{(3.26)})
admit a momentum along the $Y^5$ direction.
In terms of the corresponding supergravity solutions this fact
translates into to the possibility of adding a plane wave along the
common direction to all constituent branes. The resulting solutions
provide new embeddings of the $D=5$ black hole with non-zero
entropy. Consider, as an example, the supergravity solution (\ref{(3.20)})
with a plane wave along the $y_5$ direction. By calculating the  area of the 
horizon (in the Einstein frame) we obtain the following expression for
the Bekenstein-Hawking entropy
\eqn
S_{BH}=\frac{A_3}{4G^{5}_N}\sqrt{\cos^2{\z}\ \a_1\a_2\a_3}
=\frac{A_3}{4G^{5}_N}\sqrt{|Q_{134}P_2Q_3|}\ ,
\label{(3.29)}\eeqn
where the charges $Q_{134}$ and $P_2$ are defined in  (\ref{(3.22)}) and 
(\ref{(3.23)}),
and $Q_3$ is the electric charge associated with the KK modes in the
$y_5$ direction. We note that the entropy is generally non-vanishing
(it vanishes for the special case $\cos{\z}=0$) and is in a non-trivial way
related to the D-2-brane direction in the $Y^1Y^2$ plane, as well 
as to the D-4-brane world-volume gauge field (see (\ref{(3.25)})).

$D=4$
black holes with finite entropy can be obtained by adding a KK
monopole  which will yield the same structure of the entropy:
\eqn
S_{BH}=\frac{A_2}{4G^{4}_N}\sqrt{\cos^2{\z}\ \a_1\a_2\a_3\a_4}
=\frac{A_2}{4G^{4}_N}\sqrt{|Q_{134}P_2Q_3P_4|}\ ,
\label{(3.30)}\eeqn
where $P_4$ is the magnetic charge associated with the KK monopole. 
The quantisation constraints on $\z$  and $\alpha_i$ are  expected to ensure that
 the entropies (\ref{(3.29)}) and (\ref{(3.30)}) take the  form of the type:
$S_{BH}=2\pi\sqrt{q_{123}p_2q_3}$ and $S_{BH}=2\pi\sqrt{q_{123}p_2q_3p_4}$, respectively, where ($q_{123},\ q_3,\ p_2,\ p_4$) take integer values.

In order to make contact with the D-brane configuration used in the
black hole entropy counting in \cite{CallMald} we perform a T-duality
transformation along $Y^2$ on the D-brane configuration (\ref{(3.25)}). The
corresponding supergravity solution interpolates between the $1_D\perp
5_D$ and the D-3-brane solutions. For the sake of comparison with
(\ref{(2.20)}) we arrange this new D-brane configuration in the following way
\eqn
\arr{rl}
3:&Y^1,Y^4,Y^5\ ,\\
  &F_{45}=\cot{\z}\ ,\\
5:&Y^1,Y^2,Y^3,Y^4,Y^5\ ,\\
    &F_{23}=-\tan{\z}\ ,
\earr
\label{(3.31)}\eeqn
where the D-branes carry momentum along the $Y^1$ direction. It is
interesting that this D-brane configuration and the D-brane
configuration  (\ref{(2.20)}) differ only by their world-volume 
2-form gauge field strengths. 
However, their interpretation as non-threshold intersections
of D-branes is quite different. The Bekenstein-Hawking entropy (\ref{(3.29)})
is now written as
\eqn
S_{BH}=\frac{A_3}{4G^{5}_N}\sqrt{\cos^2{\z}\ \a_1\a_2\a_3}
=\frac{A_3}{4G^{5}_N}\sqrt{|Q_1P_2Q_3|}\ ,
\label{new1}\eeqn
where $Q_1$ is the charge associated with the D-string that is bounded to
the D-3-brane, $P_2$ is the D-5-brane charge and $Q_3$ the charge associated
with the KK modes along the $Y^1$ direction.

We remark that this entropy does depend on the angle parameter
$\z$; its value {\it decreases} as this parameter increases, and it approaches
zero as this parameter approaches $\pi /2$.
The microscopic picture in 
terms of the D-brane configuration is described by 
non-zero values of the world-volume
gauge fields in (\ref{(3.31)}), which {\it decrease} the contribution to the
microscopic  entropy and render the entropy zero
as $\cos{\z}\to 0$ (In this limit the world-volume 2-form gauge field 
strengths in (\ref{(3.31)}) approach zero and infinity, respectively.). 
  In conclusion, the new parameter $\z$ associated with the D-brane 
world-volume gauge fields, probes the microscopic structure of these  
more general black holes.

\section{Conclusions}

In this paper we have constructed new supergravity $p$-brane solutions
of superstring theory and M-theory. This was done by using the $SL(2,\bR)$
electro-magnetic duality of the $D=8$, $N=2$ supergravity or
the dimensional reduction along a boost on configurations of two 
orthogonally intersecting M-branes. In particular, we have shown 
that there is a multitude of new BPS D-brane bound states with 
non-vanishing worldvolume gauge fields. Specifically, we have
concentrated on configurations of two $p$-branes intersecting at an angle different from $\pi/2$. 

The main motivation for our work was to provide new D-brane
configurations that are relevant for black hole physics. In fact,
the microscopic origin of the  black hole entropy in terms of
D-brane configurations, when the corresponding world-volume gauge fields
form a condensate [as in the cases (\ref{(2.20)}) and  (\ref{(3.31)})],
opens an interesting avenue in the study of black hole physics
in string theory. In particular, black holes which are obtained upon compactification of  D-brane configurations  intersecting at an angle, have the entropy  (e.g., (\ref{(3.29)}) and (\ref{(3.30)})) which  explicitly depends on this angle. The same  (quantised) parameter in turn specifies the  world-volume gauge field condensates (\ref{(3.31)}).

Another interesting problem is to study the relationship between the stringy
description of these new D-brane configurations and their world-volume
gauge theories. The appearance of many new features  is expected, as it is  already  the case for the special cases considered in \cite{HashTayl}.

\section*{Acknowledgments}
We are grateful to  K. Chan, F. Larsen and G. Papadopoulos for helpful discussions. The work is  supported by   JNICT
(Portugal) under programme PRAXIS XXI (M.S.C.),  the U.S. DOE Grant
No. DOE-EY-76-02-3071 (M.C.), the National Science Foundation Career
Advancement Award No. PHY95-12732 (M.C.) and the NATO collaborative
research grant CGR No. 940870 (M.C.).

\end{document}